# Characterization of hydrogenated amorphous silicon sensors on polyimide flexible substrate

M. Menichelli[1,*], L.Antognini[2], S.Aziz[3], A. Bashiri[4,5], M.Bizzarri[1,6], L.Calcagnile[3], M. Caprai[1], D. Caputo[7,8], A.P. Caricato[3], R. Catalano[9], D.Chilà[10,11], G.A.P. Cirrone[9], T.Croci[1,12], G. Cuttone[9], G. De Cesare[7,8], S.Dunand[2], M.Fabi[10,13], L.Frontini[14], C.Grimani[10,13], M. Ionica[1], K. Kanxheri[1,6], M. Large[4], V.Liberali[14], N.Lovecchio[7,8], M.Martino[3], G. Maruccio[3], G.Mazza[15], A. G. Monteduro[3], A. Morozzi[1], F. Moscatelli[1,16], A. Nascetti[7,17], S. Pallotta[10,11], A. Papi[1], D. Passeri[1,12] *Senior Member IEEE*, M.Pedio[1,16], M. Petasecca[4] *Member, IEEE*, G.Petringa[9], F.Peverini[1,6], L.Piccolo[15], P.Placidi[1,12], G.Quarta[3], S. Rizzato[3], G. Rossi[1,6], F.Sabbatini[10,13], L. Servoli[1], A. Stabile[14], C. Talamonti[10,11], J. E. Thomet[2], L. Tosti[1], M.Villani[10,13], R.J. Wheadon[15], N. Wyrsch[2], N.Zema[1,18].

*Abstract*— Hydrogenated amorphous silicon (a-Si:H) is a material having an intrinsically high radiation hardness that can be deposited on flexible substrates like Polyimide. For these properties a-Si:H can be used for the production of flexible sensors. a-Si:H sensors can be successfully utilized in dosimetry, beam monitoring for particle physics (x-ray, electron, gamma-ray and proton detection) and radiotherapy, radiation flux measurement for space applications (study of solar energetic particles and stellar events) and neutron flux measurements. In this paper we have studied the dosimetric x-ray response of n-i-p diodes deposited on Polyimide. We measured the linearity of the photocurrent response to x-rays versus dose-rate from which we have extracted the dosimetric x-ray sensitivity at various bias voltages. In particular low bias voltage operation has been studied to assess the high energy efficiency of these kind of sensor. A measurement of stability of x-ray response versus time has been shown. The effect of detectors annealing has been studied. Operation under bending at various bending radii is also shown.

[1.] INFN, Sez. di Perugia, via Pascoli s.n.c. 06123 Perugia (ITALY)
[2.] Ecole Polytechnique Fédérale de Lausanne (EPFL), Institute of Electrical and Microengineering (IME), Rue de la Maladière 71b, 2000 Neuchâtel, (SWITZERLAND).
[3.] INFN and Dipartimento di Fisica e Matematica dell'Università del salento, Via per Arnesano, 73100 Lecce (ITALY)
[4.] Centre for Medical Radiation Physics, University of Wollongong, Northfields Ave Wollongong NSW 2522, (AUSTRALIA)
[5.] Najran University, King Abdulaziz Rd, Najran. (Saudi Arabia)
[6.] Dip. di Fisica e Geologia dell'Università degli Studi di Perugia, via Pascoli s.n.c. 06123 Perugia (ITALY)
[7.] INFN Sezione di Roma 1, Piazzale Aldo Moro 2, Roma (ITALY)
[8.] Dipartimento Ingegneria dell'Informazione, Elettronica e Telecomunicazioni, dell'Università degli studi di Roma via Eudossiana, 18 00184 Roma (ITALY).
[9.] INFN Laboratori Nazionali del Sud, Via S.Sofia  62, 95123 Catania (ITALY)
[10.] INFN Sez.  di Firenze, Via Sansone 1, 50019 Sesto Fiorentino (FI) (ITALY)
[11.] Department of Experimental and Clinical Biomedical Sciences "Mario Serio"- University of Florence Viale Morgagni 50, 50135 Firenze (FI) (ITALY).
[12.] Dip. di Ingegneria dell'Università degli studi di Perugia, via G.Duranti 06125 Perugia (ITALY)
[13.] DiSPeA, Università di Urbino Carlo Bo, 61029 Urbino  (PU) (ITALY)
[14.] INFN Sezione di Milano Via Celoria 16, 20133 Milano (ITALY)
[15.] INFN Sez. di Torino Via Pietro Giuria, 1 10125 Torino  (ITALY)
[16] CNR-IOM, via Pascoli s.n.c. 06123 Perugia (ITALY)
[17.] Scuola di Ingegneria Aerospaziale Università degli studi di Roma. Via Salaria 851/881, 00138 Roma. (ITALY).
[18.] CNR Istituto struttura della Materia, Via Fosso del Cavaliere 100, Roma (ITALY).
* Corresponding author, Mauro.menichelli@pg.infn.it

## I. Introduction

HYDROGENATED amorphous silicon (a-Si:H) is a disordered semiconductor that can be deposited by plasma-enhanced chemical vapor deposition (PECVD) from a mixture of Silane ($SiH_4$) and hydrogen at typical temperatures of 180-250 °C. Plasma excitation for the material used to fabricate the sensors described in this paper, is performed using VHF at 70 MHz [1]. Due to the low deposition temperature of a-Si:H it can be easily deposited on flexible materials like Polyimide (PI). The disordered nature of a-Si:H, also includes the presence of dangling bonds. In pure amorphous silicon, these dangling bonds, lead to a highly defective material. However, the passivation process by hydrogenation allows the reduction, by several orders of magnitude, of the density of the defects and also increases the bandgap. This results in making a-Si:H a viable material for radiation detector fabrication, for solar cells production and also for the development of electronic devices [2,3].

Another relevant feature of a-Si:H is its excellent radiation resistance. Photons [4,5], protons [6] and recently also neutrons [7] radiation tests have been performed on a-Si:H solar cells and detectors. The results of photon and proton tests are summarized in [8].

The HASPIDE project [9] is devoted to the development of a-Si:H sensors deposited on PI having either a n-i-p diode structure or charge selective contact device structure [10]. The main application foreseen for these sensors includes: beam monitoring for high-energy physics applications and for clinical beams, and as TRansmission Detectors (TRDs) both for electron beams for radiotherapy and proton accelerators for hadron therapy. Additional interesting fields of application for these sensors include x-ray beam dose profiling for medical and industrial applications [11], detectors for solar flare events monitoring to be used in space missions [12], and neutron detection for industrial, nuclear safeguard and homeland security.

In this paper we report the x-ray response of 5 mm x 5 mm and 2 mm x 2 mm n-i-p diodes both having 2.5 µm thickness. Both size of diodes have been studied for leakage current versus biasing voltage. Also the photocurrent at various dose rates has been studied in order to extract the radiation sensitivity of devices at various bias voltages including very low voltages (0-1 V). Preliminary annealing effects and long-term (about 6 h) stability of response measurements have been performed and the operation of bent detectors has also been tested.

## II. X-RAY RESPONSE OF N-I-P DIODES ON PI

The radiation sensors for the HASPIDE project have two different configurations n-i-p diodes and charge selective contact devices (CSC). N-i-p diodes are formed by a thin (e.g. tens of nm) layer of p-doped a-Si:H, a thicker (1-10 µm) layer of intrinsic (undoped) a-Si:H and a thin layer of n-type doped a-Si:H. Charge selective contact devices [10] are based on a three-layer structure featuring a thin layer of metal-oxides where with a small activation energy (such as $TiO_2$), a thick layer of intrinsic a-Si:H, and a thin layer of metal-oxides with a large activation energy (such as $MoO_x$ or $WO_x$). In this paper tests on n-i-p diodes will be shown. The detailed structure of this device is shown in Fig. 1. On the top of a 25 µm thick PI substrate an Aluminum layer is deposited via sputtering (90 nm thickness). In order to avoid diffusion of Aluminum in a-Si:H, a layer of 5 nm of Chromium is deposited over the Aluminum using the same technique. On top of this metal a layer of n-doped a-Si:H is deposited (ca. 20 nm thickness) via PECVD from a mixture of $SiH_4$, $H_2$ and $PH_3$. To create a n-i-p structure a 2.5 µm layer of intrinsic a-Si:H is deposited on top of the n-doped layer (via PECVD). On top of this layer a patterned deposition of p-doped a-Si:H is performed (PECVD of a mixture of $SiH_4$, $H_2$ and $B_2H_4$). On the p-doped pattern a deposition of Indium Tin Oxide (ITO) is performed via sputtering; a top view of the resulting detector is shown in Fig.2.

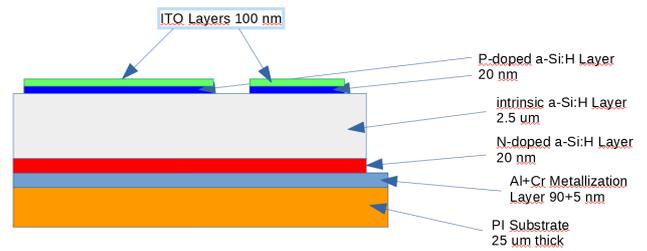

Fig 1 Layout of HASPIDE n-i-p diode prototype.

The x-ray setup used for the measurements described in this paper is shown in Fig.3. The irradiated sample includes five 2 mm x 2 mm and one 5 mm x 5 mm n-i-p diodes and is glued and bonded (using a copper based conductive glue) to a PCB frame. This is connected to an interface board in connection with a Keithley 2400 SMU (Source Measuring Unit) that is used for biasing the sensor and measuring the output current with a resolution of about 1 pA. The sensor is exposed to x-rays generated by a 10 W x-ray tube from Newton Scientific operating at 50 kV maximum voltage and 200 µA maximum current [13].

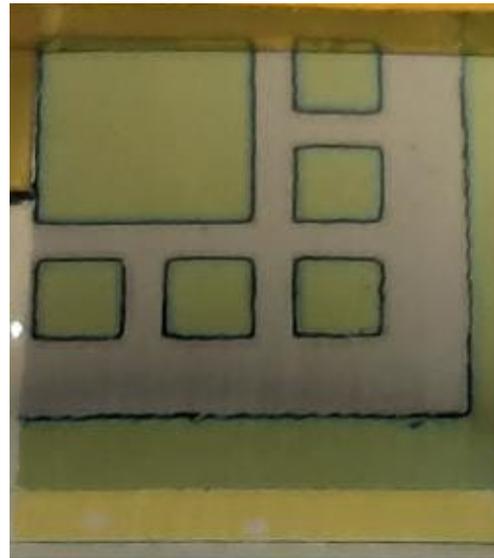

Fig.2 Picture from the top of the detector array before packaging. One 5 mm x 5 mm device and five 2 mm x 2 mm devices are deposited on PI. The light green area is the ITO contacts, the grey area is the intrinsic a-Si:H and the grey area below the green-yellow area is the Cr+Al back contact.

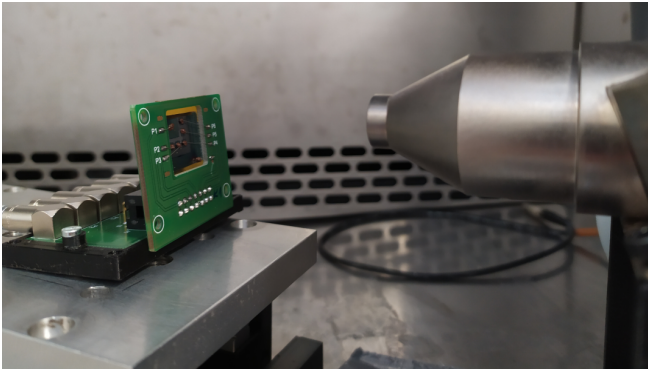

Fig.3 Setup for X-ray testing. The detector shown in Fig.2 is connected to the SMU through a PCB interface frame. The picture also shows the x-ray tube collimator. The entire setup is enclosed in a climatic chamber for temperature stabilization.

Fig.4 displays the measurements of dark current at room temperature versus voltage for a 2 mm x 2 mm (small diode) and for a 5 mm x 5 mm (large diode) sensor. The power absorption at 1 V of the detector is 10 pW (large diode) and 1 pW (small diode) while at 10 V bias is below 10 nW for a large diode and 1 nW for a small diode. The ratio between the leakage currents of the large and the small diode is approximately equal to the ratio of the sensor areas.

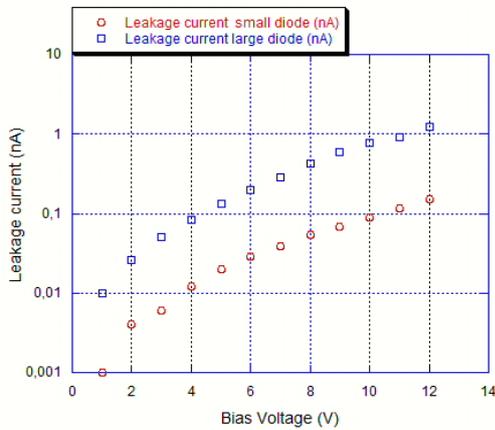

Fig.4 Leakage current at room temperature versus bias voltage for the 2 mm x 2 mm device (small diode) and for the 5 mm x 5 mm device (large diode).

In order to measure dosimetric sensitivity, the detectors have been irradiated with x-rays using a tube voltage of 40 kV in the range from about 20 to 200 µA of tube current. The dose rate of the emitted radiation in this setup was measured according to the procedure shown in [7]. The large and the small sensors were irradiated in the dose rate range from 0.36 to 3.11 cGy/s and the photocurrents have been measured at different values of the detector bias. After the subtraction of the leakage current, the photocurrent has been plotted versus x-ray tube emitted dose rate at various bias voltages and the results are plotted in Fig.5 for the large diode and in Fig. 6 for one small diode. From these figures it is possible to infer the very good linearity of the detector responses in the measured range.

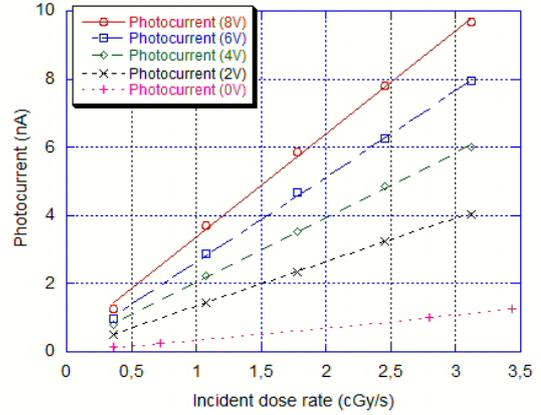

Fig.5 Net photocurrent versus incident x-ray dose rate for the large (5 mm x 5 mm) device at various bias voltages.

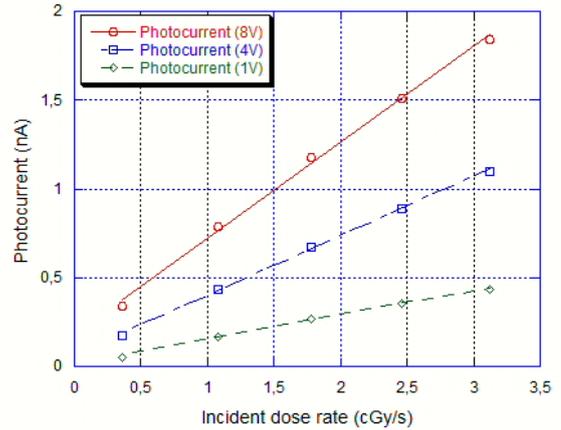

Fig.6 Net photocurrent versus incident x-ray dose rate for the small (2 mm x 2 mm) device at various bias voltages.

The dosimetric sensitivities and linear regression coefficients have been extracted from the slopes of the lines coming from the linear fit; these calculated quantities are shown in Table I.

TABLE I. DOSIMETRIC SENSITIVITY VALUES AND LINEAR REGRESSION COEFFICIENTS

| Device Area ($mm^2$) | Bias Voltage | Dosimetric sensitivity (nC/cGy) | Regression coefficient R |
|---|---|---|---|
| 5 x 5 | 0V | 0.367 | 0.99999 |
| | 2V | 1.283 | 0.99991 |
| | 4V | 1.900 | 0.99975 |
| | 6V | 2.505 | 0.99972 |
| | 8V | 3.027 | 0.99926 |
| 2 x 2 | 1V | 0.137 | 0.99878 |
| | 4V | 0.335 | 0.99961 |
| | 8V | 0.540 | 0.99881 |

From these measurements, it is possible to determine the power consumption under irradiation. For the small diode at 1V bias, the absorbed power is ranging from 54 pW at 0.36 cGy/s to 432 pW at 3.11 cGy/s, while at 8 V it ranges respectively from 2.74 nW to 14.76 nW. For the large diode also the photocurrent at 0V bias has been measured with negligible power consumption while at 8 V the power consumption of the detector ranges from 10.24 nW to 77.36 nW. These data demonstrate the very low power consumption of these sensors.

## III. Detector Long-term stability of n-i-p devices measured with X-rays

A test on the longer term of the x-ray response of these devices has also been performed: a 5 mm x 5 mm device has been irradiated for $2.1 \times 10^4$ s at 40 kV tube bias voltage with a dose rate of 0.4 cGy/s. Fig.7 shows the raw data of the time profile of the collected photocurrent (red data points), the slow rise of the current, in the stabilization phase, may be due to thermal effects. For this reason a background dark current, evaluated by a linear extrapolation between the dark current before and after irradiation, has been subtracted to the raw photocurrent. The result of this correction is shown with the green data points. After the application of this algorithm, we are able to compensate for the thermal effect of the x-ray irradiation.

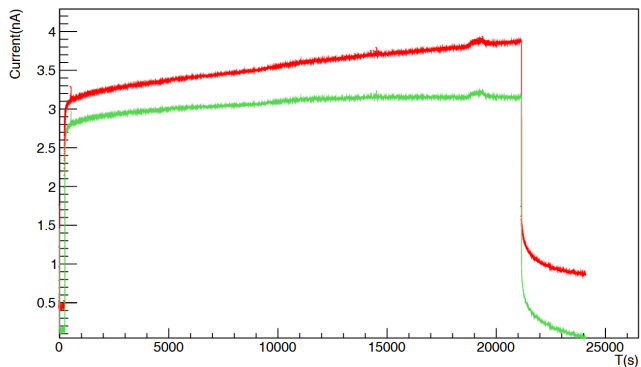

Fig.7 Raw photocurrent of a 5 mm x 5 mm n-i-p device biased at 8 V and irradiated at the rate of 0.4 cGy/s versus time (red data points) and photocurrent corrected for the increase of background leakage current due to thermal effect (green data points).

## IV. Operation of the sensor during bending

In order to test the operation of the device when bent, a bending test has been performed on the sensor using a 615 nm optical laser. The test setup is shown in Fig.8. The detector is glued on a flexible Polyimide PCB support; the support is fixed on a jig equipped with two jaws which, approaching each other, cause the curvature of the support and therefore of the sensor. A camera observes the support from the side and using an appropriate software (ImageJ [14]) it is possible to superimpose a circle on the bent support image and calculate the radius of curvature of the bending (Fig.9). By illuminating the sensor with the laser the photocurrent has been measured on the sensor, to check if there are relevant changes due to bending. The laser is mounted on a movable support in order to keep the same distance from the sensor to correct for the small divergence. The measurement started from a flat configuration of the support and the measured value was used to the photocurrent measurements taken during bending. The curvature was then increased to a bending radius of about 8 mm and then decreased back again to a flat position. The results of the relative photocurrent versus curvature radius are shown in Fig. 10a. The points on the blue line were taken during bending radius decrease and the points on the brown line were taken during a bending radius increase, in order to check for degradation or hysteresis; Fig 10b shows the percentage of the deviation from the initial photocurrent.

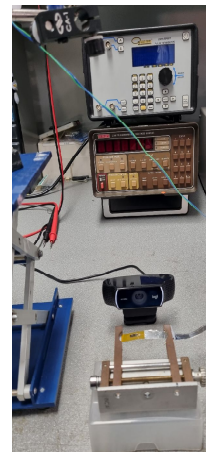

Fig.8 The setup for the bending test. The sample is glued on a kapton PCB shielded by aluminum. The support is mounted on a jig with two jaws to change the bending radius. The camera and the laser diode (on the top) are also shown.

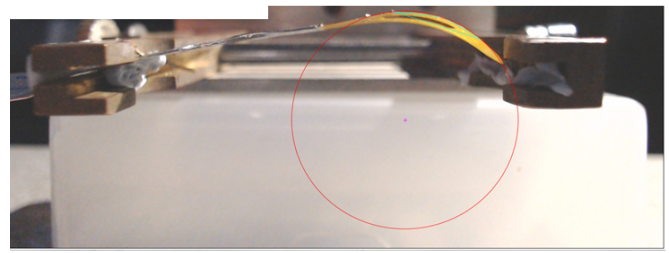

Fig.9. The bent sensor on the jig and the curvature radius measurement.

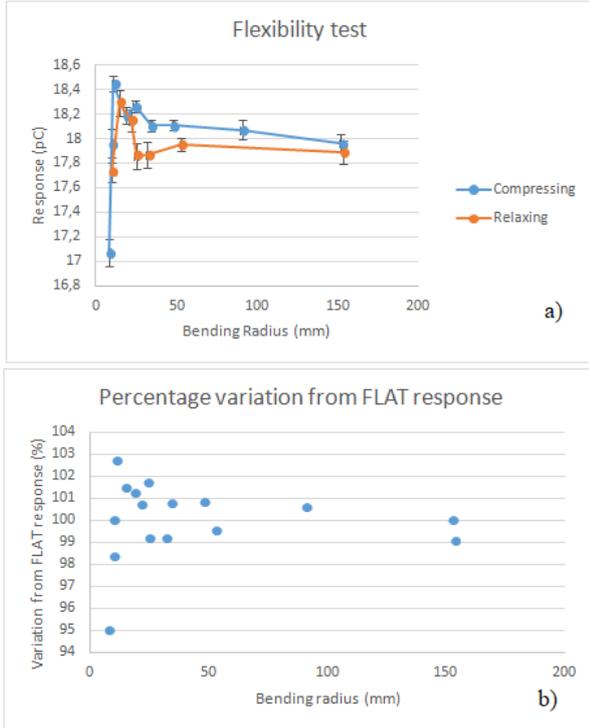

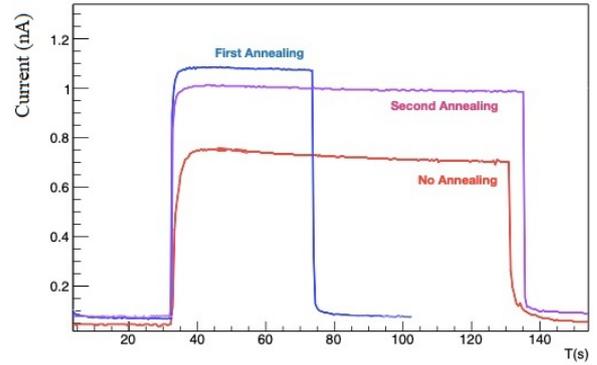

Fig.11 Photocurrent amplitude signal of an a-Si:H sensor before annealing (orange line), after 12 hours of annealing (blue line) and 24 hours annealing (purple line).

Fig. 10 Results from bending measurements. a) Charge response vs. bending radius. Measurements on the blue line are taken during compression and measurements on the brown line are taken during relaxation. b) Deviation from flat response (100%) vs bending radius.

From the measurements we can see that except for the point at 8 mm bending radius, where the photocurrent is 95% of the flat response, the deviation from the flat response is below 3% and there is only a little difference between response during curvature radius decrease (compression) and response during curvature radius increase (relaxation).

## V. ANNEALING STUDIES

During radiation damage tests with neutrons [7] especially at $10^{16}$ $n_{eq}/cm^2$ for n-i-p devices, a very large recovery effect due to annealing has been observed. The irradiated components, after the annealing, improved their characteristics not only in comparison with the post-irradiation phase but also in comparison to the performances measured before irradiation. For these reasons we tested non-irradiated components before and after the annealing. The annealing was performed in two phases: a) 12-hours of baking at 100 °C and b) 24-hours (overall) of baking at 100 °C. Fig. 11 shows the photocurrent versus time, where at an irradiation of 2.456 mGy/s, we can notice the time response of the various stages of the annealing test. Fig. 12 shows the photocurrent versus x-ray dose rate for the component after the first and second phase of the annealing test. From this graph we can notice a large increase in the dosimetric sensitivity response (from 1.8 to 18.0 nC/cGy) after the first phase while after the second phase a small decrease of the response is observed (from 18.0 to 17.4 nC/cGy).

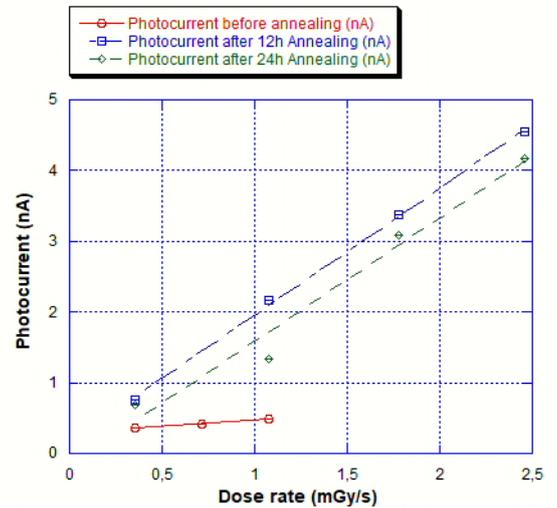

Fig.12 Photocurrent versus dose rate response of a fresh detector (red line and dots), 12 hr annealed (blue dot and line) and 24 hr annealed device (green dot and line).

From this test we can infer that annealing can be beneficial not only on irradiated components but also on non-irradiated ones. Although the best duration for this annealing is still under optimization, these results suggest it will be below 12 hours.

## VI. CONCLUSIONS

Two different size (2 mm x 2 mm and 5 mm x 5 mm ) a-Si:H n-i-p devices on PI, (having 2.5 µm thickness) have been built in the context of the HASPIDE project, aiming at the construction of flexible planar detectors for radiation flux measurements and neutron detection. These devices have been tested for leakage current, dosimetric sensitivity at various bias voltages, long-term behavior in response, flexibility and annealing. The results show a very good linearity in the dose rate range tested, in addition to a good sensitivity and leakage current scaling with the area of the devices. Power requirements of the detectors ranges from tens of pW to tens of nW depending on sensor size and bias voltage. Bias voltage is related to the dosimetric sensitivity of the device; the greater the needed sensitivity, the higher bias voltage required and therefore a higher power consumption is expected. Long term behavior of the sensor is sufficiently stable especially if power dissipation is correctly implemented. Flexibility under operation is very good; above 1 cm of bending radius the photocurrent variations are contained within ± 3% of the flat-sensor response. Furthermore, we observe beneficial effect on the device annealed for 12h at the temperature of 100 °C.

ACKNOWLEDGEMENTS

The HASPIDE project is funded by INFN through the CSN5 and was partially supported by the "Fondazione Cassa di Risparmio di Perugia" RISAI project n. 2019.0245. F. Peverini has a PhD scholarship funded by the PON program. M. J. Large is supported by the Australian Government Research Training Program (AGRTP) scholarship and the Australian Institute of Nuclear Science (AINSE) Post-Graduate Research Award (PGRA). A. Bashiri is sponsored by Najran University, Saudi Arabia. L. Antognini and J. E. Thomet are supported by the Swiss National Science Foundation (grant number 200021_212208/1).